\documentclass{skaox2006}
\def\grtsim{\mathrel{\hbox{\rlap{\hbox{\lower2pt\hbox{$\sim$}}}\raise2pt\hbox{$>$}}}} 
\def\lesssim{\mathrel{\hbox{\rlap{\hbox{\lower2pt\hbox{$\sim$}}}\raise2pt\hbox{$<$}}}}

\def\degree{\nobreak\ifmmode{^\circ}\else{$^\circ$}\fi}

\newcommand{\whzsr}{W~Hz$^{-1}$~sr$^{-1}$}

\newcommand{\lbol}{$L_{\rm bol}$}

\def\msol{M$_{\odot}$}

\def\amean{$\langle  \hat{a} (z)   \rangle$}
\def\sp{$\hat{a}$}

\newcommand{\mbh}{$m_{\bullet}$}

\begin{document}

   \title{The cosmic spin of the most massive black holes}

   \author{Alejo Mart\'\i nez-Sansigre\inst{1}\inst{2}\inst{3}
          \and
          Steve Rawlings\inst{2}
          }

   \institute{Institute of Cosmology and Gravitation, University of Portsmouth, Dennis Sciama Building, Burnaby Road, Portsmouth, PO1 3FX, United Kingdom
         \and
           Astrophysics, Department of Physics, University of Oxford, Keble Road, Oxford OX1 3RH, United Kingdom
\and
SEP{\it net}, South-East Physics network
             }

   \abstract{Under the assumption that jets in active galactic nuclei
     are powered by accretion and the spin of the central supermassive
     black hole, we are able to reproduce the radio luminosity
     functions of high- and low-excitation galaxies. High-excitation
     galaxies are explained as high-accretion rate but very low spin
     objects, while low-excitation galaxies have low accretion rates
     and bimodal spin distributions, with approximately half of the
     population having maximal spins.  At higher redshifts ($z\sim$1),
     the prevalence of high accretion rate objects means the typical
     spin was lower, while  in the present day Universe is dominated by low accretion rate
     objects, with bimodal spin distributions.  } \maketitle
%
%________________________________________________________________
%
\section{Introduction}

Active galactic nuclei (AGN) are galaxies with signs of non-stellar
activity in their centres, believed to be powered by supermassive
black holes (SMBHs). AGN produce jets which are observable at radio
frequencies. For a given
accretion rate, AGN can produce jets which vary in radio luminosity by
several orders of magnitude. Conversely, for a given radio luminosity,
some AGN are found to have very high accretion rate, while others are
found to have very low ones.

The hidden variable behind this `radio loudness' of AGN has often been
connected with the spin of the SMBH, \sp, and this is known as the
`spin paradigm' (e.g. Wilson \& Colbert 1995; Sikora et
al. 2007). Indeed theory and simulations suggest that spinning SMBHs
should be able to power considerable jets (e.g. Blandford \& Znajek,
1977, Hawley \& Krolik 2006, Tchekhovskoy et al. 2010), and indeed
observations of the most powerful radio galaxies require an extremely
efficient mechanism for the production of jets (e.g. Punsly 2007,
Fernandes et al. 2011).

Recently, Fender et al. (2010) have noted a lack of correlation
between the reported measurements of jet powers and black hole spin
for galactic black holes in X-ray binaries. As we discuss in detail in
Mart\'\i nez-Sansigre \& Rawlings (2011, hereafter MSR11), the
uncertainties in both measurements are very large, so that the lack of
observed correlation does not provide robust evidence against the spin
paradigm (see Section~7.2 of MSR11 for a detailed discussion). 

In this work we infer the cosmic distribution of SMBHs spins under the
assumption that spin is the major variable explaining the huge
variations in jet power.  Indeed models and simulations of spinning
black holes can provide enough variation in jet power to explain most
of the variation in radio loudness of AGN (see Section 3 of MSR11).

Under this assumption, we derive the spin distributions for SMBHs with
high and low accretion rates, and infer the cosmic spin
history. Throughout our work we concentrate on the most massive black
holes, with \mbh$\geq10^{8}$~\msol. This is achieved by working only with radio luminosity densities $\geq10^{23}$ \whzsr, where all the AGN black hole masses are known to be  $\geq10^{8}$~\msol\, (McLure et al. 2004, {Smol{\v c}i{\'c}} et al. 2009).

  \begin{figure*}
   \centering
   \vspace{140pt}
  \includegraphics{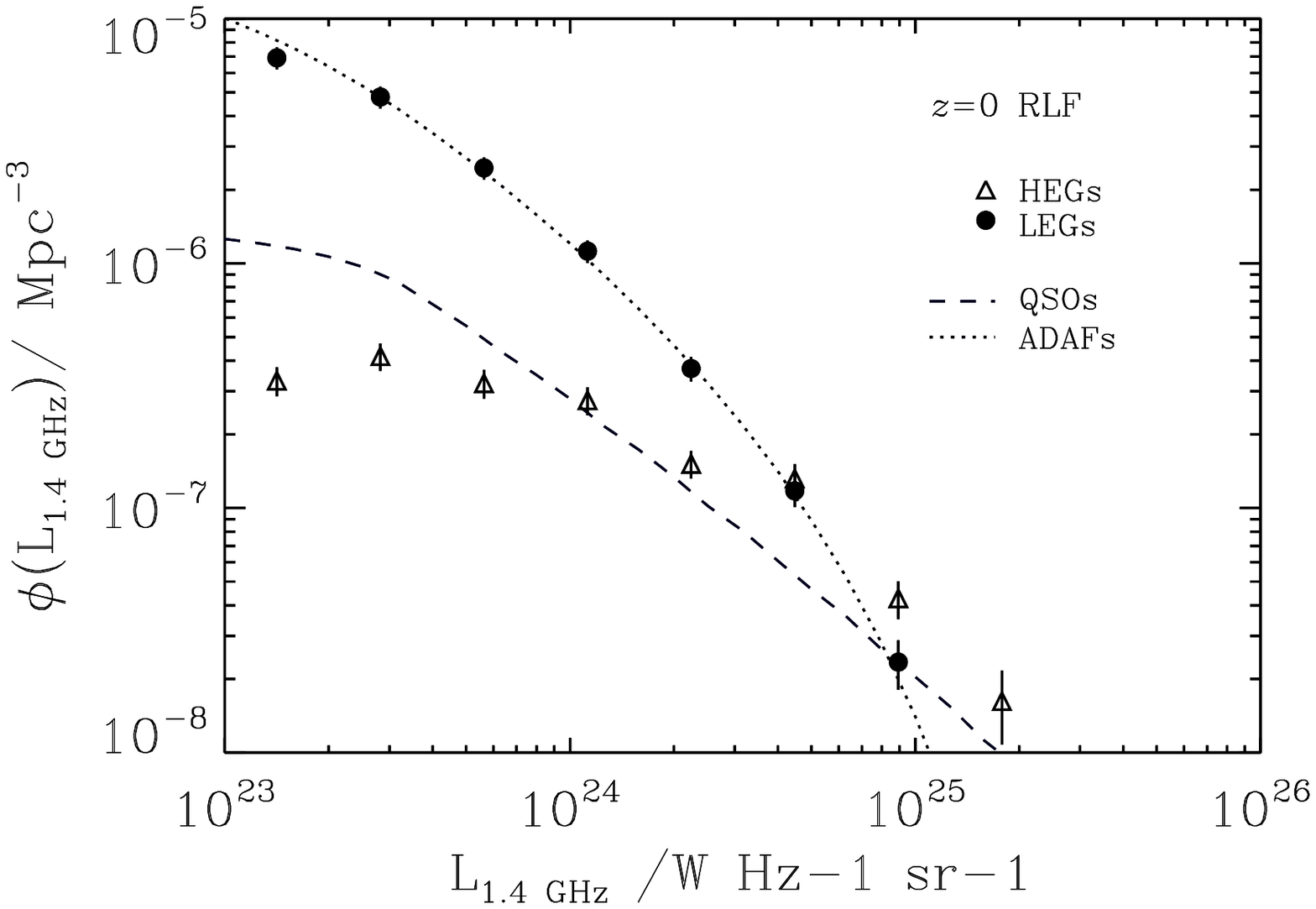}
  \includegraphics{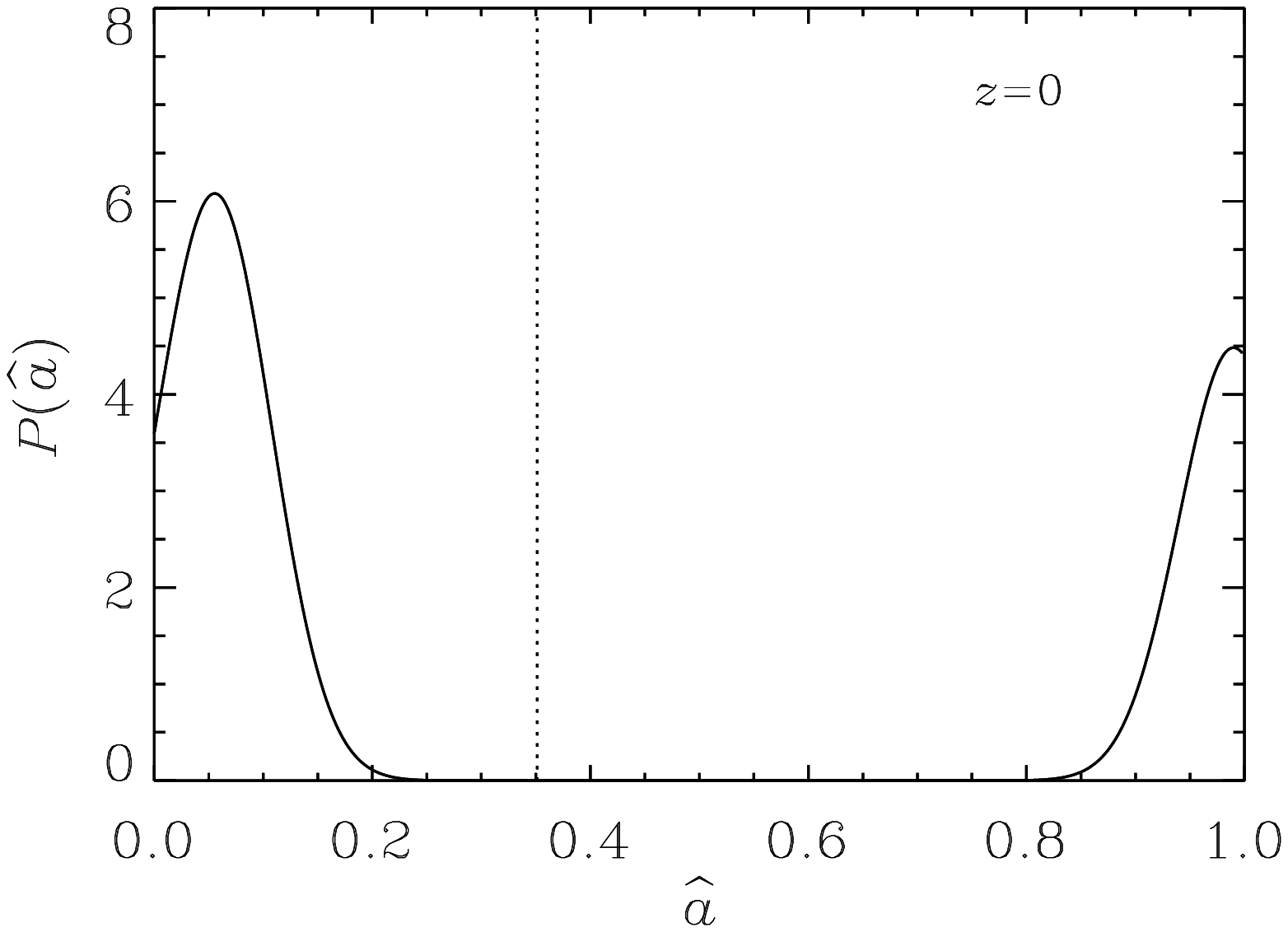}
\vspace{12pt}
   \caption{ (Left panel): Best-fit to the local radio LF of the ADAF
     and QSO components. The black circles represent the observed
     radio LF of LEGs, the empty triangles represent the HEGs. (Right
     panel): Spin distribution for all SMBHs in the locl Universe,
     obtained by combining the individual spin distributions of ADAFs
     and QSOs, weighted by their space densities (see
     Equation~\ref{eq:dbn_smbh}).
            \label{fig:z0_lf}
           }
    \end{figure*}

\subsection{Assumptions}

The bolometric power available for radiation is given by:

\begin{equation}
L_{\rm bol} = \epsilon(\hat{a}) \dot{m}_{\bullet} c^{2} 
\label{eq:lbol}
\end{equation}

 \noindent where $L_{\rm bol}$ is the bolometric luminosity due to
 radiation, $\dot{m}_{\bullet}$ is the rate of accretion of mass onto
 the SMBH, and the term $\epsilon(\hat{a})$ is the radiative efficiency (Novikov \& Thorne 1973).

We assume that the power available for the production
of jets can be described as a function of accretion rate and spin only:

\begin{equation}
Q_{\rm jet} = \eta(\hat{a}) \dot{m}_{\bullet} c^{2}
\label{eq:qjet}
\end{equation}

\noindent where $Q_{\rm jet}$ is the jet power and $\eta(\hat{a})$ is the jet
efficiency.  This is a simplification, which ignores the dependence of
the jet power on the geometry of the accretion flow. For more details
see MSR11. In these proceedings we only show the results using one
particular set of jet efficiencies, from the 3D general relativistic
magnetohydrodynamic simulations of Hawley \& Krolik (2006). However,
in MSR11 we show the results for a set of six different efficiencies,
and find the results to be robust.

Finally, to convert from jet power to observed ratio luminosity at 151 MHz, we follow the conversion derived by Willott et al. (1999):

\begin{equation}
\left( { Q_{\rm jet} \over {\rm W}}\right)= 3\times10^{38}
f^{3\over2}\left( {L_{\nu 151} \over 10^{28} ~{\rm W Hz^{-1}
    sr^{-1}}}\right)^{6\over 7}
\label{eq:will}
\end{equation}

\noindent and assume $f=$20 (see MSR11 for more details, and also
Cavagnolo et al. 2010). The term $f$ is one of the dominant sources of uncertainty in our work. We convert from 151~MHz to 1.4~GHz assuming a
power law with index -0.75.

   \begin{figure*}
   \centering
   \vspace{140pt}
  \includegraphics{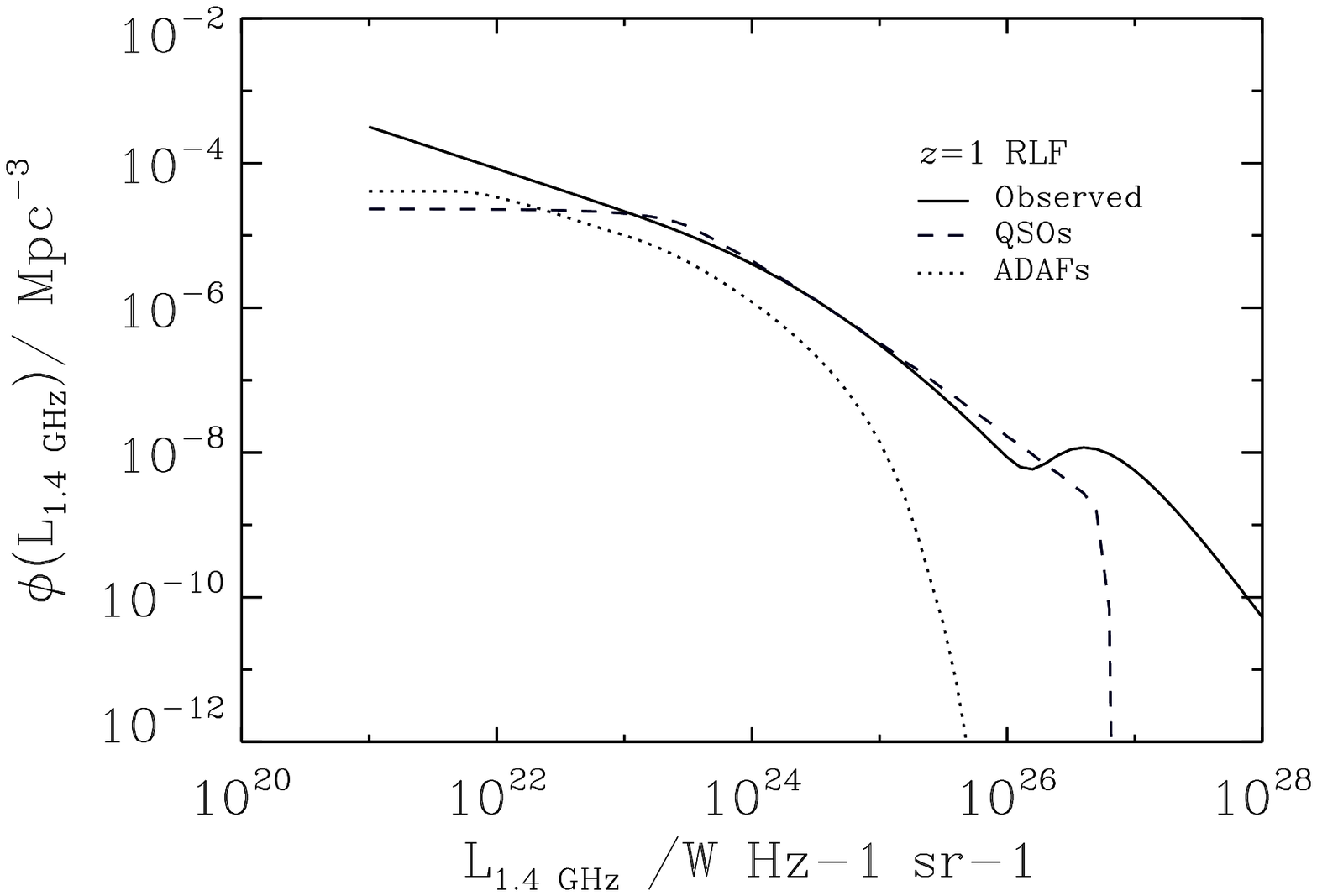}
  \includegraphics{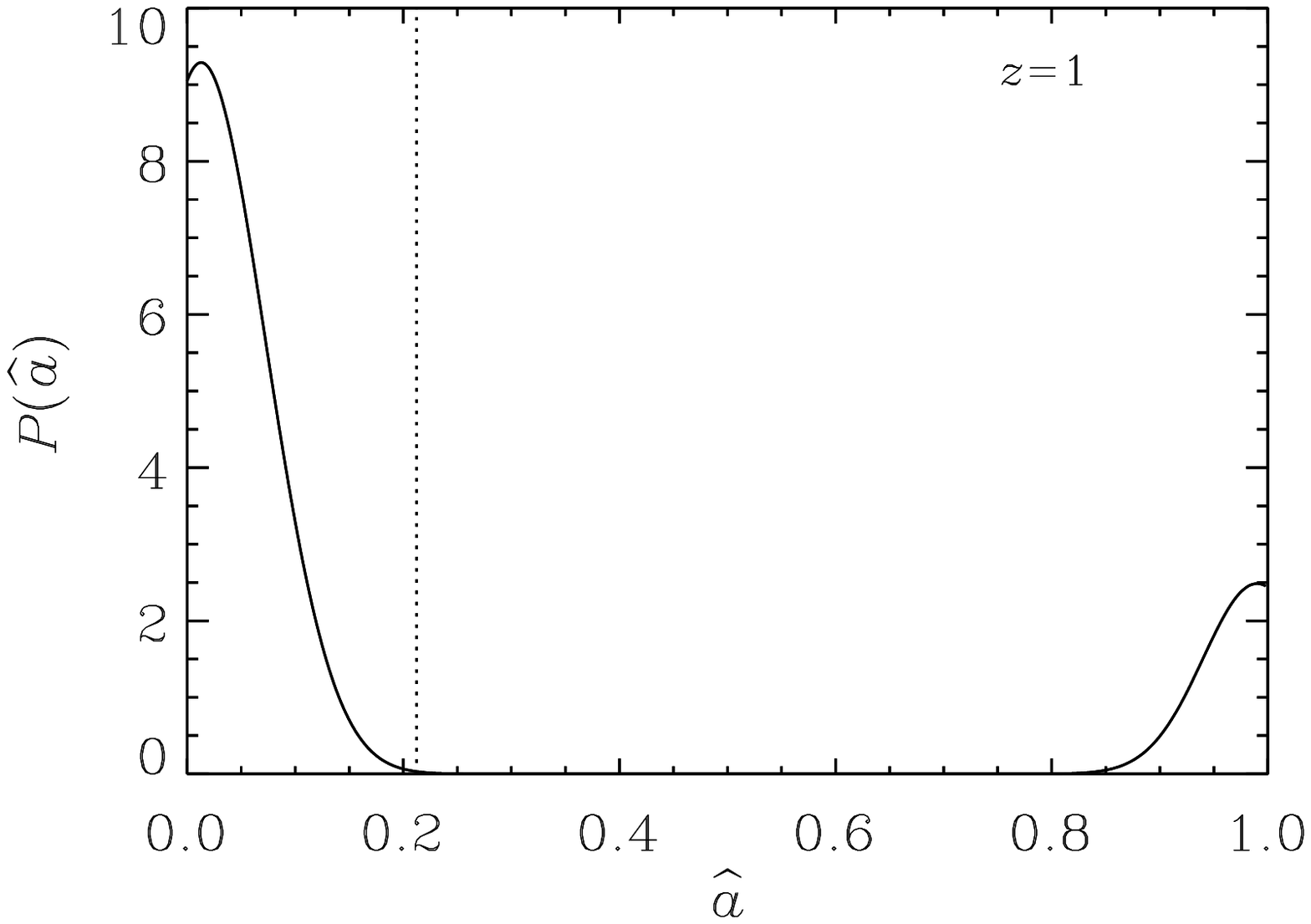}
\vspace{12pt}
   \caption{(Left panel): Comparison of the $z=$1 predicted and
     observed radio LFs. The solid line represents the observed
     luminosity function (Willott et al. 2001, Smol{\v c}i{\'c} et
     al. 2009). The dashed line is the predicted QSO contribution, the dotted line is the ADAF contribution. (Right panel): The corresponding spin distribution of SMBHs at $z=$1. 
            \label{fig:z1_lf}
           }
    \end{figure*}

\section{Modelling the local radio luminosity function}

Best et al. (in prep.) have classified the radio sources of the local
radio luminosity function (LF), according to their optical
spectra. Sources have been classified into high-excitation and
low-excitation galaxies (HEGs and LEGs, respectively), and individual
radio LFs have been derived.

We model these two populations independently. For the HEGs, we assume
that these represent SMBHs accreting at a significant fraction of
their Eddington limiting (`QSOs'). The X-ray LF
provides us with the space density of SMBHs with high accretion rates
(Silverman et al. 2008). The bolometric luminosity, \lbol, can be
estimated from the X-ray luminosity $L_{\rm X}$, via a bolometric
correction, $C_{\rm X}$, so that \lbol$=$$C_{\rm X}L_{\rm X}$. Using
equations~\ref{eq:lbol} and \ref{eq:qjet}, the jet power can be
estimated using:

\begin{equation}
Q_{\rm jet} = {\eta(\hat{a}) C_{\rm X}L_{\rm X} \over  \epsilon(\hat{a})}.
\end{equation}

Hence,  given a  spin we can infer the jet power from the X-ray luminosity. We model the space density of HEGS with a given radio luminosity as:

\begin{eqnarray}
\phi_{\rm QSO}(L_{\nu})
=  {{\rm d}Q_{\rm jet}\over {\rm d}L_{\nu}}\int P_{\rm QSO}(\hat{a})\phi( L_{\rm X})  {{\rm d} L_{\rm X} \over {\rm d}\dot{m}}  {\rm d}\hat{a} .
\label{eq:phi_qso}
\end{eqnarray}

The term $\phi_{\rm QSO}(L_{\nu})$ is the modelled radio LF of QSOs, which we
use to explain the HEGs, $\phi( L_{\rm X})$ is the X-ray LF, and
$P_{\rm QSO}(\hat{a})$ is the distribution of spins for the
QSOs. ${\rm d}Q_{\rm jet}/ {\rm d}L_{\nu}$ is computed from
Equation~\ref{eq:will}.  For details of the derivation of
Equation~\ref{eq:phi_qso} we refer the reader to Section 4.1 of
MSR11. The X-ray LF is integrated above X-ray luminosities that can only be reached by SMBHs with \mbh$\grtsim10^{8}$~\msol. We note that we include a correction factor to account for the
fraction of QSOs that are optically-thick to Compton scattering and
hence missed by the hard X-ray surveys (e.g. Mart\'\i nez-Sansigre et
al. 2007).

For the LEGs, we model these as low-Eddington rate objects
(`ADAFs'). We do not have direct access to an ADAF LF. However, the
local mass function of SMBHs is known (e.g. Graham et al. 2007), so we
can use it to model the space density of ADAFs. Given a black hole
mass, \mbh, the Eddington limiting accretion rate is given by
$\dot{m}_{\rm E}c^{2} = L_{\rm E} m_{\bullet} / \epsilon$, where
$L_{\rm E}$ is the Eddington limiting luminosity, 1.3$\times10^{31}$ W
\msol$^{-1}$. Hence, given a black hole mass and an Eddington rate,
$\lambda\equiv \dot{m}/\dot{m}_{\rm E}$, we can estimate the jet
power:

\begin{equation}
Q_{\rm jet} = {\eta(\hat{a}) \lambda \dot{m}_{\rm E}}.
\end{equation}

Given the space density of SMBHs with a mass \mbh, $\phi(m_{\bullet})$, the space density of radio sources powered by ADAFs, $\phi_{\rm ADAF}(L_{\nu})$,  is given by:

\begin{eqnarray}
\phi_{\rm ADAF}(L_{\nu}) = \nonumber \\
{{\rm d}Q_{\rm jet}\over {\rm d}L_{\nu}} \int P_{\rm ADAF}(\hat{a})\int P(\lambda )  \phi(m_{\bullet}) {{\rm d} \lambda \over {\rm d} \dot{m}'} {\rm d}m_{\bullet} {\rm d}\hat{a}, 
\end{eqnarray}

\noindent where $P_{\rm ADAF}(\hat{a})$ is the spin distribution of
ADAFs and $P(\lambda )$ is our prior for the distribution of Eddington
ratios (again see MSR11 for more details). Given our ignorance of the distribution of Eddington ratios, we assign a flat prior in log space, with uniform probability density in the range $-8 \leq {\rm log}_{10}\lambda\leq -2$.

The only free parameters are the terms describing the spin
distributions $P_{\rm QSO}(\hat{a})$ and $P_{\rm ADAF}(\hat{a})$. We
use the data to constrain the best fitting parameters for three
different spin distributions: a power law, a single gaussian and a
double gaussian. The bayesian odds ratio is used to choose
between models.

Figure~\ref{fig:z0_lf} (left panel) shows the observed radio LFs for
HEGs and LEGs, and overlayed are the best-fitting $\phi_{\rm
  QSO}(L_{\nu})$ and $\phi_{\rm ADAF}(L_{\nu})$.  The best fitting
distribution for the QSOs is a single gaussian centred around
\sp$=$0.00, for the ADAFs it is a double gaussian, centred at
\sp$=$0.06 and 0.99, and with the high-spin gaussian having an
amplitude of 0.79 compared to the low-spin gaussian.

Hence, we find that the high-accretion rate SMBHs (QSOs/HEGs) have typically low spins, while amongst the low-accretion rate SMBHs (ADAFs/LEGs) there is  a high fraction of objects with high spins. 

The typical spin distribution for all SMBHs at  $z$$=$0  can be estimated from the weighted mean:

\begin{eqnarray}
P_{\rm SMBH}(\hat{a}) = \nonumber \\
{   \sum \phi_{\rm QSO}(L_{\nu}) P_{\rm QSO}(\hat{a}) + \sum \phi_{\rm ADAF}(L_{\nu}) P_{\rm ADAF}(\hat{a})  \over \sum \phi_{\rm QSO}(L_{\nu}) + \sum \phi_{\rm ADAF}(L_{\nu})  }
\label{eq:dbn_smbh}
\end{eqnarray}

The right panel of Figure~\ref{fig:z0_lf} shows $P_{\rm SMBH}(\hat{a})$ in the local Universe. 

\section{Predicting the $z=$1 radio luminosity function}

We now test whether our modelling can reproduce the radio LF at
$z=$1. From the observed X-ray LF, we can model the evolution of the
QSOs. 

No such information is available for the ADAFs, however. Given that
the bulk of the growth of the most massive black holes, with
\mbh$\geq10^{8}$~\msol, ocurred in the range $1 \grtsim z \grtsim 3$,
make the assumption that there has been negligible evolution in the
mass function of SMBHs. This assumption is supported by the weak
evolution observed amongst radio sources of moderate luminosity
(e.g. Smol{\v c}i{\'c} et al. 2009).  Hence, at $z=$1 we use the local
SMBH mass function, with the same distribution of Eddington ratios.

Figure~\ref{fig:z1_lf} (left panel) shows the resulting radio LF for
QSOs (dashed) and ADAFs (dotted), compared to the observed radio LF
(solid line). Without any more free parameters, only the assumption of
no evolution for the ADAFs, we are able to approximately reproduce the
$z=$1 radio LF over six decades of radio luminosity.

\begin{figure*}
   \centering
   \vspace{140pt}
  \includegraphics{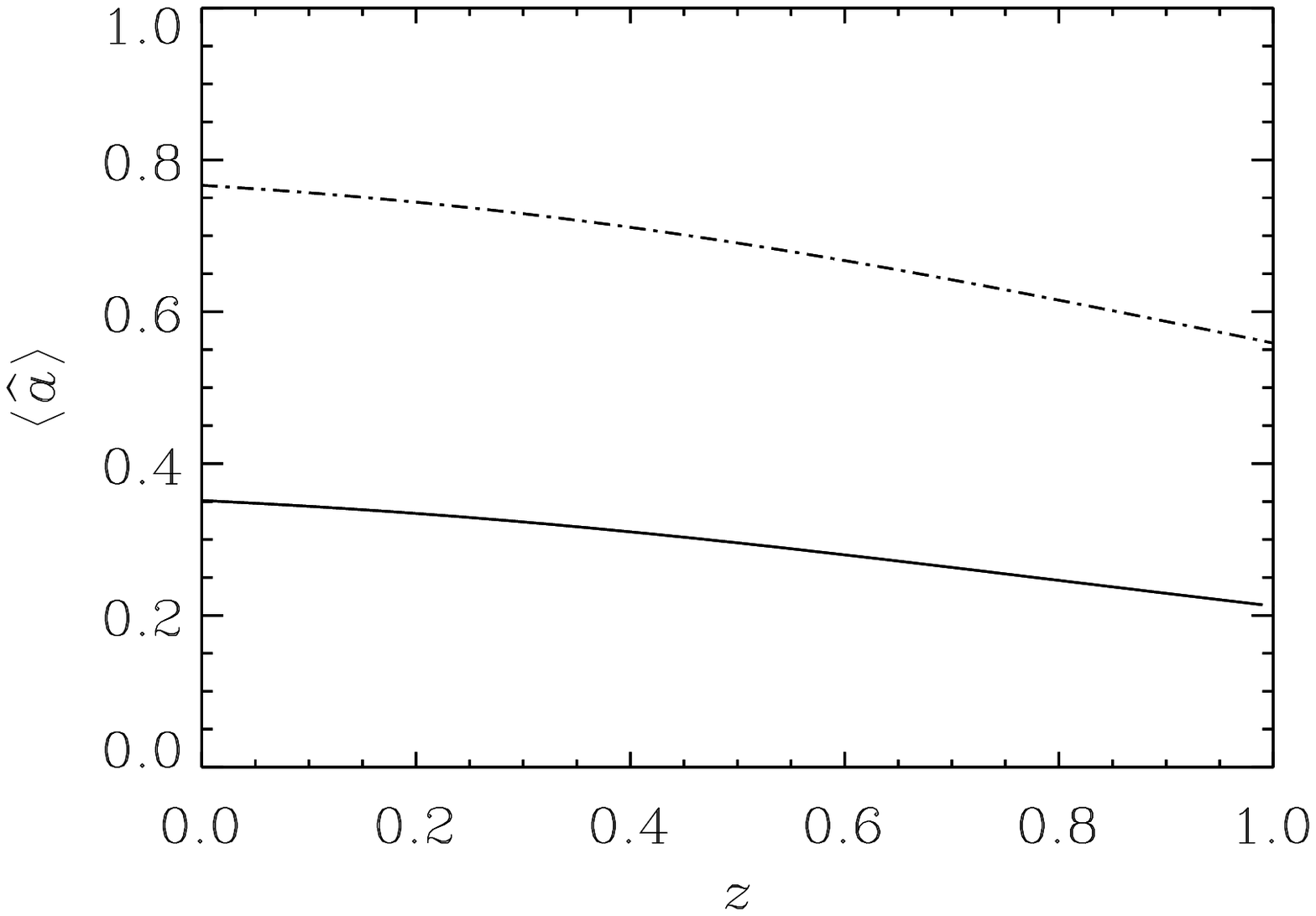}
 \includegraphics{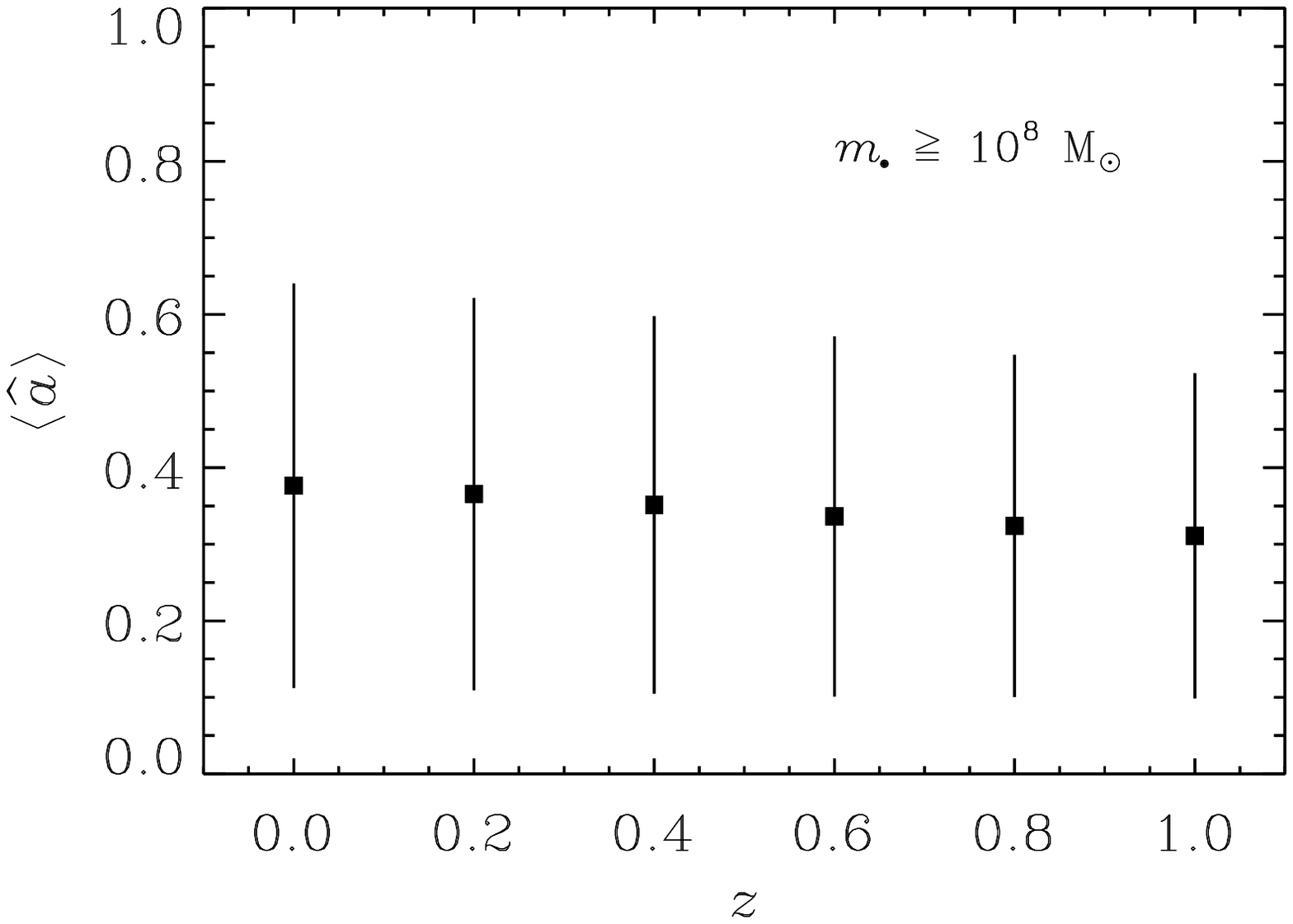}
\vspace{12pt}
   \caption{(Left panel): Evolution of the mean spin inferred from our spin distributions for all SMBHs as a function of $z$. The solid line is the mean, the dashed-dotted line shows the standard deviation (note the -1$\sigma$ goes along the x axis). (Right panel): Evolution of the mean spin from the cosmological simulation of Fanidakis et al. (2011), for \mbh$\geq10^{8}$~\msol. 
            \label{fig:az}
           }
    \end{figure*}

\section{The evolution in cosmic spin of SMBHs}

Applying Equation~\ref{eq:dbn_smbh} to $z=$1, we obtain the spin
distribution for all SMBHs at $z=$1, which is shown in the right panel
of Figure~\ref{fig:z1_lf}. We see that at high redshift, the fraction
of SMBHs with high spin is lower than at low redshift.

It is expected that at $z\geq1$ the mass function of SMBHs will begin
to decrease significantly in space density.  This suggests that the
space density of ADAFs will also decrease, so that we expect the spin
distributions at higher redshifts to be even more dominated by the \sp$\sim$0
term.

The evolution of the spin of SMBHs can therefore be understood as a
gradual switch between two populations. At high redshift, a
high-accretion rate but low spin population dominates (the QSOs). This
population decreases strongly in space density as $z$ decreases,
revealing a second population of low-accretion sources, which have a
bimodal spin distribution (the ADAFs).
 
This evolution is best described by the spin distributions, as
illustrated by the right-hand panels of Figures~\ref{fig:z0_lf} and
\ref{fig:z1_lf}. However, for completeness we also show the evolution
of the mean spin (the expectation value) as a function of redshift, in
Figure~\ref{fig:az} (left panel).  This shows a modest evolution from
\amean$\sim$0.2 at $z\sim1$, to \amean$\sim$0.35 in the local
Universe, and when extrapolated to higher redshifts, predicts a mean
radiative efficiency $\sim$0.065, in excellent agreement with
observational constraints (e.g. Mart\'\i nez-Sansigre \& Taylor 2009).

\section{Discussion}

We have found that the population dominanting the radio luminosity
function switches from high-accretion rate objects with low spins at
high redshift, to low-accretion rate objects with a bimodal
distribution of spins, with approximately half the population having
maximal spin. 

This can be explained by the effects on the spin of SMBHs by the two
mechanisms by which these grow: accretion and mergers. Continous
accretion along one plane would lead to SMBHs having essentially
maximal spin. However, it has been recently suggested that SMBHs do
not accrete in this way, but are rather subject to 'chaotic
accretion', where matter falls in from different
directions. Approximately half of the matter has angular momentum in
the opposite direction to the SMBH, and the final spin is therefore,
on average, close to 0. Hence accretion provides a mechanism for
spinning SMBHs down (King et al. 2008).

Major mergers of SMBHs, on the other hand, are likely to increase the
final spin (e.g. Rezzolla et al. 2008). When two black holes of similar mass merge, the angular
momentum of the final orbit will be significant compared to the final
mass, so that the spin of the coalesced black hole will be large.

The evolution of the spin distributions inferred in our work is in
good agreement with a picture where at high redshift the spins are
typically low due to accretion. Major mergers will occur both at high
and low redshift, but at high redshift there is a larger supply of
cold gas available (Obreschkow \& Rawlings 2009), so that accretion
will spin the SMBHs back down.

At low redshift, the cold gas is running out, so after the mergers the
breaking mechanism is no longer present. If a large fraction of the
SMBHs have undergone a recent major merger, then a large fraction of
the SMBHs will retain a high spin.

%Our results agree quantitatively as well as qualitatively with this
%picture. The SMBHs with \mbh$\geq10^{8}$~\msol\, are typically hosted
%by massive galaxies, with stellar masses $\grtsim10^{11}$~\msol\, and
%these are estimated to have undergone, on average, 0.7 mergers since
%$z\sim1.2$ (Robaina et al. 2010). ...

A detailed cosmological simulation of the growth of SMBHs was
performed by Fanidakis et al. (2011), which included the effects of
black hole mergers as well as chaotic accretion. The left panel of
Figure~\ref{fig:az} shows the mean spin from their simulation, as a
function of redshift. The simulation shows an almost identical
evolution to that inferred from our work, and a very similar, large
variance.

\begin{acknowledgements}
 We  thank Philip Best and Nikolaos Fanidakis for sharing their
 data.  A.M.-S. gratefully acknowledges a PDF
 from the UK STFC, reference ST/G004420/1. This effort was partly
 supported by the EC FP6, SKADS.
\end{acknowledgements}


\begin{thebibliography}{}


\bibitem[\protect\citeauthoryear{{Blandford} \& {Znajek}}{{Blandford} \&
  {Znajek}}{1977}]{1977MNRAS.179..433B}
{Blandford} R.~D.,  {Znajek} R.~L.,  1977, \mnras, 179, 433

\bibitem[\protect\citeauthoryear{{Cavagnolo}, {McNamara}, {Nulsen}, {Carilli},
  {Jones} \& {B{\^i}rzan}}{{Cavagnolo} et~al.}{2010}]{2010ApJ...720.1066C}
{Cavagnolo} K.~W.,  {McNamara} B.~R.,  {Nulsen} P.~E.~J.,  {Carilli} C.~L.,
  {Jones} C.,    {B{\^i}rzan} L.,  2010, \apj, 720, 1066


\bibitem[\protect\citeauthoryear{{Fanidakis}, {Baugh}, {Benson}, {Bower},
  {Cole}, {Done} \& {Frenk}}{{Fanidakis} et~al.}{2011}]{2011MNRAS.410...53F}
{Fanidakis} N.,  {Baugh} C.~M.,  {Benson} A.~J.,  {Bower} R.~G.,  {Cole} S.,
  {Done} C.,    {Frenk} C.~S.,  2011, \mnras, 410, 53



\bibitem[\protect\citeauthoryear{{Fender}, {Gallo} \& {Russell}}{{Fender}
  et~al.}{2010}]{2010MNRAS.406.1425F}
{Fender} R.~P.,  {Gallo} E.,    {Russell} D.,  2010, \mnras, 406, 1425


\bibitem[\protect\citeauthoryear{{Fernandes}, {Jarvis}, {Rawlings},
  {Mart{\'{\i}}nez-Sansigre}, {Hatziminaoglou}, {Lacy}, {Page}, {Stevens} \&
  {Vardoulaki}}{{Fernandes} et~al.}{2011}]{2011MNRAS.411.1909F}
{Fernandes} C.~A.~C.,  {et~al.},  2011, \mnras, 411, 1909



\bibitem[\protect\citeauthoryear{{Graham}, {Driver}, {Allen} \& {Liske}}{{Graham}
  et~al.}{2007}]{2007MNRAS.378..198G}
{Graham} A.W.,  {Driver} S.P.,    {Allen} P.D.,  2007, \mnras, 378, 198

\bibitem[\protect\citeauthoryear{{Hawley} \& {Krolik}}{{Hawley} \&
  {Krolik}}{2006}]{2006ApJ...641..103H}
{Hawley} J.~F.,  {Krolik} J.~H.,  2006, \apj, 641, 103


\bibitem[\protect\citeauthoryear{{King}, {Pringle} \& {Hofmann}}{{King}
  et~al.}{2008}]{2008MNRAS.385.1621K}
{King} A.~R.,  {Pringle} J.~E.,    {Hofmann} J.~A.,  2008, \mnras, 385, 1621



\bibitem[\protect\citeauthoryear{{Mart{\'{\i}}nez-Sansigre} \&
  {Rawlings}}{{Mart{\'{\i}}nez-Sansigre} \& {Rawlings}}{2011}]{2011arXiv1102.2228M}
{Mart{\'{\i}}nez-Sansigre} A.,  {Rawlings} S.,  2011, \mnras, 414, 1937 

\bibitem[\protect\citeauthoryear{{Mart{\'{\i}}nez-Sansigre}, {Rawlings},
  {Bonfield}, {Mateos}, {Simpson}, {Watson}, {Almaini}, {Foucaud}, {Sekiguchi}
  \& {Ueda}}{{Mart{\'{\i}}nez-Sansigre} et~al.}{2007}]{2007MNRAS.379L...6M}
{Mart{\'{\i}}nez-Sansigre} A.,  et~al.,  2007, \mnras, 379, L6

\bibitem[\protect\citeauthoryear{{Mart{\'{\i}}nez-Sansigre} \&
  {Taylor}}{{Mart{\'{\i}}nez-Sansigre} \& {Taylor}}{2009}]{2009ApJ...692..964M}
{Mart{\'{\i}}nez-Sansigre} A.,  {Taylor} A.~M.,  2009, \apj, 692, 964


\bibitem[\protect\citeauthoryear{{McLure}, {Willott}, {Jarvis}, {Rawlings},
  {Hill}, {Mitchell}, {Dunlop} \& {Wold}}{{McLure}
  et~al.}{2004}]{2004MNRAS.351..347M}
{McLure} R.~J.,  {Willott} C.~J.,  {Jarvis} M.~J.,  {Rawlings} S.,  {Hill}
  G.~J.,  {Mitchell} E.,  {Dunlop} J.~S.,    {Wold} M.,  2004, \mnras, 351, 347


\bibitem[\protect\citeauthoryear{{Novikov} \& {Thorne}}{{Novikov} \&
  {Thorne}}{1973}]{1973blho.conf..343N}
{Novikov} I.~D.,  {Thorne} K.~S.,  1973, in Black Holes, ed. C. Dewitt, \& B.~S. Dewitt
(New York: Gordon and Breach), 343


\bibitem[\protect\citeauthoryear{{Obreschkow} \& {Rawlings}}{{Obreschkow} \&
  {Rawlings}}{2009}]{2009ApJ...696L.129O}
{Obreschkow} D.,  {Rawlings} S.,  2009, \apjl, 696, L129


\bibitem[\protect\citeauthoryear{{Punsly}}{{Punsly}}{2007}]{2007MNRAS.374L..10P}
{Punsly} B.,  2007, \mnras, 374, L10


\bibitem[\protect\citeauthoryear{{Rezzolla}, {Barausse}, {Dorband}, {Pollney},
  {Reisswig}, {Seiler} \& {Husa}}{{Rezzolla}
  et~al.}{2008}]{2008PhRvD..78d4002R}
{Rezzolla} L.,  {Barausse} E.,  {Dorband} E.~N.,  {Pollney} D.,  {Reisswig} C.,
   {Seiler} J.,    {Husa} S.,  2008, \prd, 78, 044002


\bibitem[\protect\citeauthoryear{{Sikora}, {Stawarz} \& {Lasota}}{{Sikora}
  et~al.}{2007}]{2007ApJ...658..815S}
{Sikora} M.,  {Stawarz} {\L}.,    {Lasota} J.-P.,  2007, \apj, 658, 815

\bibitem[\protect\citeauthoryear{{Silverman}, {Green}, {Barkhouse}, {Kim},
  {Kim}, {Wilkes}, {Cameron}, {Hasinger}, {Jannuzi}, {Smith}, {Smith} \&
  {Tananbaum}}{{Silverman} et~al.}{2008}]{2008ApJ...679..118S}
{Silverman} J.~D.,  et~al.,  2008, \apj, 679, 118

\bibitem[\protect\citeauthoryear{{Smol{\v c}i{\'c}} et~al.,}{{Smol{\v c}i{\'c}}
   et~al.}{2009}]{2009ApJ...696...24S}
{Smol{\v c}i{\'c}} V.,  et~al., 2009, \apj, 696, 24

\bibitem[\protect\citeauthoryear{{Tchekhovskoy}, {Narayan} \&
  {McKinney}}{{Tchekhovskoy} et~al.}{2010}]{2010ApJ...711...50T}
{Tchekhovskoy} A.,  {Narayan} R.,    {McKinney} J.~C.,  2010, \apj, 711, 50


\bibitem[\protect\citeauthoryear{{Willott}, {Rawlings}, {Blundell} \&
  {Lacy}}{{Willott} et~al.}{1999}]{1999MNRAS.309.1017W}
{Willott} C.~J.,  {Rawlings} S.,  {Blundell} K.~M.,    {Lacy} M.,  1999,
  \mnras, 309, 1017

\bibitem[\protect\citeauthoryear{{Willott}, {Rawlings}, {Blundell}, {Lacy} \&
  {Eales}}{{Willott} et~al.}{2001}]{2001MNRAS.322..536W}
{Willott} C.~J.,  {Rawlings} S.,  {Blundell} K.~M.,  {Lacy} M.,    {Eales}
  S.~A.,  2001, \mnras, 322, 536

\bibitem[\protect\citeauthoryear{{Wilson} \& {Colbert}}{{Wilson} \&
  {Colbert}}{1995}]{1995ApJ...438...62W}
{Wilson} A.~S.,  {Colbert} E.~J.~M.,  1995, \apj, 438, 62


 
\end{thebibliography}
\end{document}